\documentclass{article}
\usepackage{srcltx}
\usepackage[english]{babel}
\usepackage{amsmath}
\usepackage{amssymb}
\usepackage{amsfonts}
\usepackage{latexsym}
\usepackage{textcomp}
\usepackage{appendix}
\usepackage{multirow}
\usepackage{booktabs}
\usepackage{rotating}
\usepackage{afterpage}
\usepackage{pdflscape}
\usepackage{url}
\usepackage{array}
\usepackage{marvosym}
\usepackage{graphics}
\usepackage{graphicx}
\usepackage{natbib}
\usepackage[table]{xcolor}
\usepackage{epsfig}
\usepackage{float}
\usepackage{authblk}

\usepackage[caption=false]{subfig}

\title{A bibliometric analysis of Bitcoin scientific production}

\author[1]{Ignasi Merediz-Sol\`a }
\author[1]{Aurelio F. Bariviera \thanks{aurelio.fernandez@urv.cat}}
\affil[1]{\scriptsize  Universitat Rovira i Virgili, Department of Business, Av. Universitat 1, 43204 Reus, Spain}

\begin{document}
\maketitle

\begin{abstract}
Blockchain technology, and more specifically Bitcoin (one of its foremost applications), have been receiving increasing attention in the scientific community. The first publications with Bitcoin as a topic, can be traced back to 2012. In spite of this short time span, the production magnitude (1162 papers) makes it necessary to make a bibliometric study in order to observe research clusters, emerging topics, and leading scholars. Our paper is aimed at studying the scientific production only around bitcoin, excluding other blockchain applications. Thus, we restricted our search to papers indexed in the Web of Science Core Collection, whose topic is ``bitcoin''. This database is suitable for such diverse disciplines such as economics, engineering, mathematics, and computer science. This bibliometric study draws the landscape of the current state and trends of Bitcoin-related research in different scientific disciplines. 
\end{abstract}

\section{Introduction \label{sec:intro}}
Bibliometric studies has become an emergent and buoyant discipline, given the importance posed on the assessment of scientific production in recent times. Eugene Garfield, with the establishment of the Institute for Scientific Information (ISI) in the 1960s, initiated the metrification of papers, journals, researchers, and institutions. Scientific papers are now compiled and indexed in large databases, which allow to measure different aspects of such papers, such as number of authors, keywords, topic, citations, institutional collaboration etc. The rationale for indexing articles is the following: authors cite other papers due to its connection with the core idea of his/her paper. Given that authors must select carefully which papers to cite, including only the most relevant and most closely related to his/her paper, most cited papers could reflect the importance of them within its discipline. Institutions can obtain valuable information about individual and aggregate impact. Therefore it could help in either faculty recruiting process or in defining the global research strategy of universities and research councils. 

However, the importance of bibliometric studies goes beyond the institutional level.  They could help new researchers of a discipline to understand the extent of a topic, emergent trends, and its evolution through time. In this sense it is different from a traditional literature survey. 

This kind of analysis is possible due to the availability of big databases such as the Web of Science. This indexing service is an important input of the evaluation process in academia. The Web of Science is a citation indexing service, administered by Clarivate Analytics, and constitutes a selective list of journals and conference proceedings, with indexing coverage from 1898. It covers more than 59 million records. The firm produces several impact metrics included in the Journal Citation Report, e.g. Impact Factor, Eigenfactor, 5-year Impact Factor, among others. These metrics are available on a subscription basis. Further details can be consulted at its website \cite{WoSinfo}. 

From a macroscopic level, we can obtain certain metrics that are common to many journals, and they are useful to different stakeholders. However, there are some features that change from discipline to discipline. There is an uneven number of researchers and journals per discipline. In recent years there has been an expansion in the number of journals and an increase in its periodicity, probably due to the expansion of the academic sector in the last decades in several countries. In addition, disciplines have different traditions regarding publication. Some disciplines, such as biomedicine, is prone to 'hyper-authorship' \citep{Cronin2001}, where a single paper includes massive collaboration, where some of them have minimal involvement in the production of the paper, and a subgroup of authors acts as reporters of the whole group.
Therefore, it is important to study the intrinsic characteristics of homogeneous disciplines and/or topics, in order to provide for a meaningful classification. 

Bitcoin research has soared in recent years. From a technological point of view, blockchain is a disrupting paradigm. It introduced the concept of distributed consensus-based validation, instead of centralized validation. Its first application, introduced by \cite{Nakamoto}, comprises the creation of a sort of means of payment, which runs parallel to the established financial system. Since its introduction, bitcoin has been increasingly become an important investment and speculative device. It got the attention of mass media, as more and more money was poured into the cryptocurrency market. At the beginning, most researchers were attracted trying to understand how blockchain worked \citep{Zyskind,Zheng2018}. Early research within economics was focused on the potential of bitcoin as a substitute of national currencies \citep{NBERw19747,Bohme2015}. Currently, most of the research from an economics and finance point of view is to analyze all its financial properties for several reasons. Firstly, because of its traded volume. Secondly, its distinct behavior \textit{vis-\`a-vis} traditional assets such as stocks, bonds or fiat currencies. Thirdly, the fact that cryptocurrencies trade on a 24/7 basis, makes studies especially interesting for algorithmic trading, and poses new challenges on how to process efficiently a continuous flux of big data. More recently, a topic which is being studied is the possibility of bitcoin to become a safe-haven asset due to its low correlation with the traditional markets \citep{Smales2018,Klein2018,Shahzad2019}, or to use it for portfolio diversification \citep{Liu2018,PLATANAKIS2019}

The aim of this paper is to analyze metadata of all the papers indexed in the Web of Science Core Collection, whose topic is ``bitcoin'', excluding other blockchain applications. This study is targeted to a broad and diverse audience. Bitcoin stakeholders come from several fields: computer programmers, investors, libertarians thinkers, and financial economists. This paper provides useful information for them on the main journals interested in publishing papers on this topic, as well as on the evolution of topics addressed in those papers. Additionally, we discuss other aspects such as highly cited papers, top publication sources and keyword analysis. 

Published research on Bitcoin begun in 2012, and until January 2019 there are 1162 papers on this topic. This paper in some aspects, extends that by \cite{Holub2018}, adding almost two year data. Unlike \cite{Holub2018}, which uses multiple data sources, we restricted our search to Web of Science. The reason for such restriction is that Web of Science is the most important database of scientific papers, whose metrics are widely used in academic assessment in several countries. Additionally, it improves the analysis adding graphical representation of citation, journal and author's networks, using VOSviewer software.

This work conducts an analytic study of the papers included in the sample using ``bibliometrix'' R package \citep{CorradoScientometrix,bibliometrix}. In addition, we add a graphical analysis of the bibliographic material using VOSviewer software \citep{vanEck2010}. This software collects the data and generates maps based on bibliographic coupling, co-authorship, citation, co-citation and co-occurrence of keywords \citep{Merigo2016}. 

The rest of the paper is structured as follows. Section \ref{sec:literature} describes the literature on bibliometric studies on econometric journals. Section \ref{sec:Data} details the data under analysis and  comments the main findings of our study. Finally, Section \ref{sec:Conclusions} draws the main conclusions.

\section{Related literature \label{sec:literature}}
Bibliometrics is a research field within library and information sciences that studies the bibliographic material by using quantitative methods \citep{Pritchard1969,Broadus1987}. Over the years, bibliometrics has become very popular for classifying bibliography and developing representative summaries of the leading results.

There are many bibliometric studies of a wide variety of issues. For example, \cite{Cobo2011} analyze the thematic evaluation of the Fuzzy Sets Theory. \cite{Bonilla2015} analyze the academic research developed in Latin America in economics between 1994 and 2013. \cite{Cancino2017614} develop a bibliometric analysis of the publications of the Computers \& Industrial Engineering between 1976 and 2015. Related to economics there are several examples. \cite{Andrikopoulos201623} performed a bibliometric analysis in the economics field by reviewing the first forty years of the Journal of Econometrics, focusing on collaboration patterns and the internationalization of research in econometrics. Another example of a bibliometric analysis of economic journals is done by \cite{Wei2019}. More specifically, \cite{Fonseca2019} conduct a bibliometric analysis of the scientific field of Behavioral Economics and Behavioral Finance, proving how they have turned into an important field of study. Furthermore, \cite{Claveau2016} introduce a combination of different bibliometric tools to the history of economics. So, combining bibliometrics with dynamic network analysis, they identified a changing specialty structure in economics from the late 1950s to 2014. In a different way, \cite{Korom2019} explored the potential of interdisciplinary perspectives by investigating the thematic overlap between economic and sociological approaches to wealth inequality.

In 2009, an anonymous individual, under the pseudonym ``Nakamoto'', published a white paper setting the grounds for the creation of a decentralized, non government controlled, currency. Those were the times of the global financial crisis, considered by many specialists to have been the most serious economic downturn since the Great Depression \citep{Almunia}. Many people stopped trusting banks. Nakamoto's idea was timely and rapidly adopted by the public. Bitcoin allows for peer-to-peer money transfer, avoiding the established financial system. In addition, transactions are encrypted. Although not anonymous, transactions are blind to national authorities. Bitcoin success has been so great, that it became tantamount to cryptocurrency. The success of bitcoin encouraged other crypto-entrepreneurs to develop their own currencies. As of February 2019, there are more than 2000 cryptocurrencies, with a total market capitalization of \$131.8 billions, and daily transactions exceeding \$24 billions \citep{coinmarketcap}. Despite the large number of cryptocoins, Bitcoin constitutes half of the market. According to \cite{Bariviera2018} most studies focus their attention on Bitcoin, rather than on the other coins. 

Unlike \cite{Miau2018}, and considering that blockchain is such a broad research area, we narrow our paper's scope only to bitcoin research. Consequently, this paper gathers all the literature produced so far regarding, exclusively, bitcoin. 

There are few previous papers on this topic. The first paper in this strand is by \cite{Liu2016}, which collects 253 articles related to bitcoin from Scopus database. This author finds three groups of papers, related to the (a) technological, (b) economic and (c) legal aspects of bitcoin. Our paper differs from \cite{Liu2016} in two aspects: (i) the origin of data and (ii) the methodology. Ours is from Web of Science, whereas Liu's is from Scopus. This is an important difference, because Scopus is a broader database, which includes other sources such as periodic magazines, which are not frequently cited in scientific journals. The second difference is regarding the treatment of data. \cite{Liu2016} uses a supervised learning algorithm, whereas we prefer to use unsupervised learning, considering the quantity and diversity of papers. The second paper in this topic is \cite{Holub2018}. They collect 4429 papers from several sources, including public working papers repositories such as ArXiv or SSRN. 
Their paper identify seven main categories of papers: (a) Technology, (b) Economics, (c) Finance, (d) Regulation, (e) Taxation, (f) Accounting, and (g) Critical Thought. The last category includes papers on political, philosophical and ethical aspects of bitcoin. Their methodology is substantially different from ours. Whereas they classify papers after reading a sample of abstracts and full articles, we use and advanced machine learning technique in order to classify papers.  In addition we study in detail authors productivity, journal citation influence, and keywords. The third paper, \cite{Corbet2019}, provides a systematic literature review on empirical economic aspects of cryptocurrencies. Their approach is based on a traditional literature review, whereas our approach is mainly a bibliometric approach. In addition, we consider a broader bibliographic \textit{corpus}, in order to provide a broader view of the scientific production around bitcoin. 

We also identified some papers dealing, tangentially, with our subject. \cite{Zeng2018} present a bibliographic analysis of the blockchain-related literature between January 2011 and September 2017, taking EI Compendex (EI) and China National Knowledge Infrastructure (CNKI) databases as the literature sources. Considering that EI is database centered in engineering literature, and CNKI provides mainly China knowledge resources, our scope and coverage is far broader. \cite{Chatterjee2018}, despite not doing a bibliometric analysis, provide a state of-the-art survey over Bitcoin related technologies and making a summary of its evolution. \cite{Dabbagh2019} present a bibliometric analysis of 995 papers dealing with blockchain. Their analysis indicate that researchers have shifted their research interests form bitcoin to blockchain in the recent two years. Their study differs from ours in their broader search query, which included, in addition to `bitcoin', the following words:  `Blockchain', `cryptocurrency', `ethereum', and `smart contract'. As a consequence, papers in their sample are more technological-focused. \cite{Yli-Huumo2016} analyzed 41 papers, excluding explicitly papers dealing with economic, legal, business and regulation perspectives of blockchain. Their study is conducted using a systematic mapping process described by \cite{Petersen2008}.

Consequently, our paper could be considered an expanded contribution to the literature, providing a full overview of the current trends on Bitcoin research, and identifying top researchers, institutions, and journals in this field.

\section{Data and results \label{sec:Data}}

This paper works with data form Web of Science Core Collection (WoS), Clarivate Analytics. We selected all indexed papers that contain 'bitcoin' as topic, which makes a total of 1162 documents, published in 703 sources (journals, books, etc.), during the period 2012-2019. These documents were (co-)authored by 2293 people. The vast majority of the documents are multi-authored, being only 322 documents single-authored. The average number of authors per document is 1.97.

Bitcoin as a research topic comprises several disciplines.  It is a recent and emerging topic, considering that the first paper is found in 2012, omitting the seminal paper by \cite{Nakamoto}. Articles cover different aspects of this novel product: legal concerns, economic perspectives or computer peculiarities. However, they are concentrated around two main research areas: computer science and business economics. Web of Science assigns indexed papers to one or more research areas. The 1162 papers considered in our sample were assigned 
1543 research areas. The top five research areas are displayed in Table \ref{tab:researchareas}

\begin{table}[htbp]
  \centering
  \caption{Main research areas assigned to papers in the sample}
    \begin{tabular}{lrr}
    \toprule
    Research Areas & Records & \% of 1543 \\
    \midrule
    Computer Science & 541   & 35\% \\
    Business Economics & 279   & 18\% \\
    Engineering & 196   & 13\% \\
    Telecommunications & 106   & 7\% \\
    Science Technology Other Topics & 79    & 5\% \\
    Total top 5 research areas & 1201  & 78\% \\
    \bottomrule
    \end{tabular}%
  \label{tab:researchareas}%
\end{table}%

The detail of yearly publications is displayed in Table \ref{tab:publicationyears}. We can observe that in the first two years of the sample, the relative increase exceeds 400\%. This is produced by the small initial figures. In more recent years the yearly increment is around 40\%. The expected scientific production for year 2019\footnote{The linear forecast for the expected number of papers is $32 \cdot 12=384$}
is 384, which gives a flat growth rate for the current year. The decreasing growth rate could signalize that research in this field is consolidating. In fact, cryptocurrencies emerged as an all-new area in science and technology almost ten years ago. During these years, the initial papers were devoted to the study of the underlying blockchain technology. Soon after it, economic and financial studies on Bitcoin begun. 

\begin{table}[htbp]
  \centering
\caption{Number of papers published by year, with 'bitcoin' as topic. Source: Web of Science Core Collection}
    \begin{tabular}{lll}
    \toprule
    \multicolumn{1}{r}{Year} & \multicolumn{1}{r}{\# articles} & \multicolumn{1}{r}{Yearly growth rate} \\
    \midrule
    \multicolumn{1}{r}{2012} & \multicolumn{1}{r}{3} & \multicolumn{1}{r}{-} \\
    \multicolumn{1}{r}{2013} & \multicolumn{1}{r}{17} & \multicolumn{1}{r}{467\%} \\
    \multicolumn{1}{r}{2014} & \multicolumn{1}{r}{90} & \multicolumn{1}{r}{429\%} \\
    \multicolumn{1}{r}{2015} & \multicolumn{1}{r}{148} & \multicolumn{1}{r}{64\%} \\
    \multicolumn{1}{r}{2016} & \multicolumn{1}{r}{192} & \multicolumn{1}{r}{30\%} \\
    \multicolumn{1}{r}{2017} & \multicolumn{1}{r}{296} & \multicolumn{1}{r}{54\%} \\
    \multicolumn{1}{r}{2018} & \multicolumn{1}{r}{384} & \multicolumn{1}{r}{30\%} \\
    \multicolumn{1}{r}{2019} & \multicolumn{1}{r}{32\textsuperscript{*}} &  \\
    \multicolumn{1}{r}{Total} & \multicolumn{1}{r}{1162} & \multicolumn{1}{r}{124\%\textsuperscript{**}} \\
    \bottomrule
          &       &  \\
    \multicolumn{3}{l}{*The projected \# of articles for 2019 is 384} \\
    \multicolumn{3}{l}{**Average growth rate per year from 2012 to 2018.} \\
    \end{tabular}%
  \label{tab:publicationyears}%
\end{table}%
%
\subsection{Corresponding author's geographical distribution}
Table \ref{tab:countriesarticles} shows that the USA is the country whose authors have published both more documents and obtained more total citations, followed by the United Kingdom in both aspects. The ten first countries accumulate the 65\% of the articles published related to Bitcoin.

Table \ref{tab:countriescitation} displays the main countries, ordered by total citations. The average number of citations per article, which is 4.2. The USA and the United Kingdom, the two countries with more articles published and total citation are above this figure, with 5.43 and 5.69 respectively. In spite of the fact that China is the third country in terms of published articles, it has the lowest average citations per article among the leading countries in total citations. It is also important to highlight that Ireland is the country with more citations per article, which can be used as a proxy for average scientific importance or quality of the articles. 

Furthermore, around 35\% of the articles published are made by authors of multiple countries. The international collaboration in the USA and India are below the average, with a multiple country publications rate of 14\% and 10\%, respectively. The international collaboration reaches a maximum for Spain, where 46\% of the papers are of this kind. 

\begin{table}
  \centering
  \caption{Top ten corresponding author's countries}
    \begin{tabular}{lrrrrr}
    \toprule
    \multicolumn{1}{c}{\multirow{2}[2]{*}{Country}} & \multicolumn{1}{c}{\multirow{2}[2]{*}{Articles}} & \multicolumn{1}{c}{\multirow{2}[2]{*}{Frequency}} & \multicolumn{1}{c}{Single country} & \multicolumn{1}{c}{Multiple country} & \multicolumn{1}{c}{\multirow{2}[2]{*}{MCP\_Ratio}} \\
          &       &       & \multicolumn{1}{c}{publications} & \multicolumn{1}{c}{publications} &  \\
    \midrule
    USA   & 249   & 23\%  & 214   & 35    & 14\% \\
    United Kingdom & 100   & 9\%   & 67    & 33    & 33\% \\
    China & 99    & 9\%   & 66    & 33    & 33\% \\
    Germany & 57    & 5\%   & 37    & 20    & 35\% \\
    Australia & 41    & 4\%   & 30    & 11    & 27\% \\
    Italy & 40    & 4\%   & 25    & 15    & 38\% \\
    India & 31    & 3\%   & 28    & 3     & 10\% \\
    Switzerland & 31    & 3\%   & 21    & 10    & 32\% \\
    France & 30    & 3\%   & 13    & 17    & 57\% \\
    Spain & 26    & 2\%   & 14    & 12    & 46\% \\
    Total 10 countries & 704   & 65\%  & 515   & 189   & 27\% \\
    \bottomrule
    \end{tabular}%
  \label{tab:countriesarticles}%
\end{table}%

\begin{table}
  \centering
  \caption{Top ten total citations per country}
    \begin{tabular}{lrr}
    \toprule
    Country & Total Citations & Average Article Citations \\
    \midrule
    USA   & 1353  & 5.4 \\
    United Kingdom & 569   & 5.7 \\
    Australia & 247   & 6.0 \\
    Germany & 222   & 3.9 \\
    Ireland & 187   & 15.6 \\
    China & 180   & 1.8 \\
    Spain & 170   & 6.5 \\
    Switzerland & 150   & 4.8 \\
    Israel & 146   & 13.3 \\
    Austria & 138   & 11.5 \\
		Total (all countries) & 5019 & 4.2\\
    \bottomrule
    \end{tabular}%
  \label{tab:countriescitation}%
\end{table}%

\subsection{Top publication sources}
Table \ref{tab:journals} shows the ten main sources publishing articles related to Bitcoin. Six of this sources are journals, and reflect the interdisciplinarity of this research field. \textit{Economics Letters} and \textit{Finance Research Letters} are two leading economics journals. \textit{Economics Letters} is, undoubtedly, the main publishing device, with 29 published paper on this topic. 
\textit{Physica A } is a physics journal, which is very friendly in publishing papers dealing with econophysics and statistical mechanics applications to economics. \textit{IEEE Access} and \textit{PLOS ONE} are two important multidisciplinary open access journals. 
Finally, \textit{Communications of the ACM} is leading publication for the computing and information technology fields, which is very much recognized among industry. 
Another important source is \textit{New Scientist}, which is a popular weekly science and technology magazine, founded in 1956.  There is also one book title ``The Digital Currency Challenge'' and published by Palgrave Macmillan US. This book details legal issues and technological developments of digital currencies in the US.
Finally, there are two conference proceedings, providing a significant number of papers. The diversity of the sources, in type and discipline, reflects the multidisciplinarity in this research topic. 

\begin{table}
  \centering
  \caption{Top ten most relevant sources}
\resizebox{\textwidth}{!}{%
    \begin{tabular}{lrr}
    \toprule
    Sources & \#Articles & Type \\
    \midrule
    Economics Letters  & 29 & Journal \\
    Physica A: Statistical Mechanics and its Applications & 19 & Journal \\
    IEEE Access & 17 & Journal \\
		PLOS ONE & 15 & Journal \\
    Finance Research Letters & 14 & Journal \\
		Communications of the ACM & 11 & Journal \\
		New Scientist & 16    & Magazine \\
		The Digital Currency Challenge & 17 & Book \\
    Financial Cryptography and Data Security FC2015 Workshop & 13 & Conference Proceedings \\
    Financial Cryptography and Data Security FC2014 Workshop & 12 & Conference Proceedings \\
    \bottomrule
    \end{tabular}%
		}
  \label{tab:journals}%
\end{table}%

\begin{table}
  \centering
  \caption{Main keywords}
    \begin{tabular}{lrlr}
    \toprule
    Author Keywords (DE)           & \# articles & Keywords-Plus (ID) & \# articles \\
    \midrule
    Bitcoin & 501   & Bitcoin & 102 \\
    Blockchain & 229   & Economics & 39 \\
    Cryptocurrency & 126   & Inefficiency & 33 \\
    Cryptocurrencies & 60    & Volatility & 30 \\
    Digital currency & 41    & Gold  & 21 \\
    Ethereum & 32    & Money & 20 \\
    Virtual currency & 31    & Internet & 18 \\
    Security & 27    & Market & 17 \\
    Money & 23    & Prices & 17 \\
    Anonymity & 20    & Returns & 17 \\
    \bottomrule
    \end{tabular}%
  \label{tab:keywords}%
\end{table}%
%
\subsection{Main keywords}
Table \ref{tab:keywords} shows the ten most used keyword in the Bitcoin articles. Web of Science provides two types of keywords: (a) Author Keywords, which are those provided by the original authors, and (b) Keywords-Plus, which  are those extracted from the titles of the cited references by Thomson Reuters (the company maintaining WoS). Keyword Plus are automatically generated by a computer algorithm.
The two more frequent Author Keywords are 'Bitcoin' and 'Blockchain'. It is remarkable that the word 'Economics' is the second most frequent Keyword-Plus, but it does not appear as an Author Keyword. It is clear that 'Economics' is too general to describe an article; authors do not used it as keyword, but the algorithm used to find Keyword-Plus does not discriminate such a useless keyword. On contrary, Keyword-Plus is precise at identifying keywords such as 'Volatility', 'Inefficiency', or 'Returns', as many economics papers are focused on these aspects of Bitcoin.  It is also relevant to notice that Blockchain is the second more used Authors Keywords as it is gaining a lot of attention among researchers but it is not in the ten more used Keywords-Plus.

\begin{table}
  \centering
\caption{Highly cited articles, descending order by number of citations.}
	\resizebox{1\textwidth}{!}{
\begin{tabular}{llll}
    \toprule
    Author(year) & Title & Source & \# Citations  \\
    \midrule
		\cite{Bohme2015} &Bitcoin: Economics, Technology, and Governance   &  Journal of Economic Perspectives  & 107   \\
		\cite{URQUHART201680}& The inefficiency of Bitcoin  &  Economics Letters  & 68     \\
    \cite{DYHRBERG201685}& Bitcoin, gold and the dollar - A GARCH volatility analysis  &  Finance Research Letters  & 58    \\
		\cite{Ciaian2016} & The economics of BitCoin price formation   &  Applied Economics  & 54     \\
		\cite{DYHRBERG2016139}&Hedging capabilities of Bitcoin. Is it the virtual gold?  &  Finance Research Letters  & 45   \\
		\cite{NADARAJAH20176}& On the inefficiency of Bitcoin   &  Economics Letters  & 41     \\
    \cite{BOURI2017192} &On the hedge and safe haven properties of Bitcoin: Is it really more than a diversifier?  &  Finance Research Letters  & 41   \\
    \cite{KATSIAMPA20173} & Volatility estimation for Bitcoin: A comparison of GARCH models   &  Economics Letters  & 38    \\
    \cite{Bariviera20171} & The inefficiency of Bitcoin revisited: A dynamic approach  &  Economics Letters  & 36    \\
		\cite{BALCILAR201774}& Can volume predict Bitcoin returns and volatility? A quantiles-based approach &  Economic Modelling  & 34    \\
    \cite{Meng} & When Intrusion Detection Meets Blockchain Technology: A Review  &  IEEE Access  & 29   \\
		\cite{URQUHART2017145} & Price clustering in Bitcoin  &  Economics Letters  & 28    \\
    \cite{KHAN2018395} & IoT security: Review, blockchain solutions, and open challenges &  Future Generation Comp. Sys.  & 24    \\
		\cite{Kuo2017} & Blockchain distributed ledger technologies for biomedical and health care applications   &  J. of the A. Med. Informatics Assoc. & 21    \\
		\cite{Zheng2018} & Blockchain challenges and opportunities: a survey   &  Int. J. of Web and Grid Services  & 16  \\
		\cite{LAHMIRI201828} & Chaos, randomness and multi-fractality in Bitcoin market   &  Chaos Solitons \& Fractals & 13   \\
		\cite{CORBET201828} & Exploring the dynamic relationships between cryptocurrencies &  Economics Letters  & 10  \\  
    \bottomrule
    \end{tabular}%
}
  \label{tab:highlycitedarticles}%
	\end{table}%

\subsection{Highly cited papers}
Table \ref{tab:highlycitedarticles} shows the list of the 17 articles categorized as a highly cited paper. According to the \cite{indicatorhandbook}, ``Highly cited papers are the top one percent in each of the 22 ESI subject areas per year. They are based on the most recent 10 years of publications. Highly Cited Papers are considered to be indicators of scientific excellence and top performance and can be used to benchmark research performance against field baselines worldwide''. This measure is useful in the sense that separates each article depending on its field and it is a known fact that depending on the field, the number of citations used per article is different. So, it is good way to highlight important articles from different fields. These papers signalize, in some way, research paths in the literature. 
\cite{Ciaian2016} study bitcoin price formation following the methodology by \cite{Barro1979}. They find that bitcoin price was mainly influenced by demand and supply (partly also by speculative investors), but that that macroeconomics has no significant effect.
\cite{URQUHART201680} develops a methodology to measure inefficiency in bitcoin using six tests. His methodology has been subsequently used in several articles. This paper concludes (and it was tested by other papers) that bitcoin was, initially, an inefficient market; but it could be in the process of moving towards a more efficient market.
\cite{DYHRBERG201685,DYHRBERG2016139} were among the first applying GARCH models to cryptocurrencies. Moreover, it was on the first articles to compare Bitcoin with Gold in order to classify bitcoin as an asset due to its characteristics. Their papers concluded that Bitcoin has a place on the financial markets and in portfolio management as it can be classified as something in between gold and the American dollar.
\cite{NADARAJAH20176} followed the methodology by \cite{URQUHART201680}, and find that a power transformation of bitcoin returns could be weakly informationally efficient.
\cite{KATSIAMPA20173} focus her attention on time series volatility, and detects short and long-run components in bitcoin conditional variance. \cite{Bariviera20171} also finds high persistence in daily variance, which makes GARCH models suitable for variance modelization. 
\cite{BOURI2017192} use a dynamic conditional correlation model in order to study the potential use of bitcoin as a safe-haven asset. Their main findings are that bitcoin is acts as a poor hedge, but it is suitable for diversification purposes.
\cite{BALCILAR201774} perform a non-parametric quantile analysis in order to analyze causal relation between trading volume and bitcoin returns and volatility. Their study reveal that volume has some predictive power on returns, but not in volatility. 
\cite{LAHMIRI201828} also detect long-range correlations in returns, and a dynamical nonlinear behavior in the time series. In addition, prices and returns exhibits multifractality probably due to the fat-tailed distributions.

\begin{table}[htbp]
  \centering
  \caption{Most productive authors}
    \begin{tabular}{llr}
    \toprule
    Authors & Institution & \# articles \\
    \midrule
    Mullan PC & Private consultant  & 17 \\
    Androulaki E & IBM Research & 10 \\
    Bouri E & University Saint-Esprit de Kaslik & 10 \\
    Roubaud D & Montpellier Business School & 10 \\
    Karame G & NEC Laboratories Europe & 9 \\
    Kiayias A & University of Edinburgh & 9 \\
    Dziembowski S & University of Warsaw & 8 \\
    Eyal I & Technion & 8 \\
    Gupta R & University of Pretoria & 8 \\
    Liu Y & National University of Defense Technology & 8 \\
    \bottomrule
    \end{tabular}%
  \label{tab:mostproductiveauthors}%
\end{table}%

Table \ref{tab:mostproductiveauthors} show the most productive authors of Bitcoin related articles. The first one is P. Carl Mullan with 17 articles published followed by Elli Androulaki, Elie Bouri and David Roubaud, all of them with 10 articles published.
%
\subsection{Degree of concentration of selected variables}
In this subsection we analyze globally several bibliometric variables, in order to show the degree of concentration of them. 
On important characteristic in bibliometric studies is the evenness of the contribution of authors, countries, and journals to a research topic. Information theory provides alternative metrics to traditional statistical measures of concentration, such as standard deviation, skewness or kurtosis. In particular, 
\cite{book:shannon1949} developed the celebrated Shannon entropy. Given a discrete distribution probability $P=\{p_j; j=1,\dots, N\}$, with $\sum_{j=1}^N p_j=1$, Shannon entropy is defined as: 
\begin{equation}
{\cal S}[P]=-\sum_{j=1}^N p_j \ln(p_j)
\label{eq:Shannon}
\end{equation}
This formula could be interpreted from different points of view. Within the data communication realm, it can be seen as the average rate of information produced by a stochastic source, and it is frequently used in data compression \citep{Huffman1952}. From a statistical mechanics point of view, it represents the degree of order/disorder of a physical system \citep{Lamberti2004119,Rosso07}. It is used in economics in order to construct measures of business concentration \citep{Horowitz1970,Hart1971}. In biological sciences, it is used as as a measure of the diversity at the species level \citep{Pielou1966,Fedor2013}. Finally, it is used in bibliometric studies in order to study the evenness/concentration distribution of different important variables such as research topics or authors, among others \citep{Polyakov2017}.

In order to make interpretation easier, it is better to normalize Shannon entropy, dividing by its maximum value. Thus, the normalized entropic concentration index reads:
\begin{equation}
{\cal H}[P]=\frac{{\cal S}[P]}{{\cal S}_{\text{max}}}=\frac{-\sum_{j=1}^N p_j \ln(p_j)}{\ln N}
\label{eq:normalized}
\end{equation}
Under this configuration $0 \leq {\cal H}\leq 1$, were 1 means that all categories are evenly represented. In other words, ${\cal H}=1$ means an absence of concentration, and ${\cal H}=0$ a concentrated distribution at one single point. 

We calculate the normalized entropic concentration index for the distribution of authors, sources, countries, research areas and citations. The results are displayed in Table \ref{tab:entropy}. We observe that papers citations are very concentrated (${\cal H}=0.3042$). In fact, only 58 out of the 1162 papers in our sample account for 50\% of all citations. 
Similarly, the research areas where Bitcoin-related papers are published are concentrated in a few areas. In fact 70\% of the papers corresponds to either computer science, business economics, engineering or telecommunications. We also detect that publications by countries is very concentrated. However, authors are rather evenly distributed. Taking together both results, it means that despite most production is concentrated in a few countries (see Table \ref{tab:countriesarticles}) authors are rather evenly distributed within those countries. Table \ref{tab:Lotka} shows that there are very few authors with 3 or more papers. Finally, we detect that there is a moderate concentration in publication sources. We identify some journals with a significant quantity and quality of publications (see Tables \ref{tab:journals} and \ref{tab:highlycitedarticles}. In particular the journals \textit{Economics Letters}, \textit{Finance Research Letters} and \textit{IEEE Access} not only are among the journals that have published many articles on Bitcoin, but also have several highly cited papers.

\begin{table}
  \centering
  \caption{Entropic concentration index (${\cal H}$) of selected variables}
    \begin{tabular}{lr}
    \toprule
    Variable & \multicolumn{1}{c}{${\cal H}$} \\
    \midrule
    Authors & 0.9432 \\
    Sources & 0.8817 \\
    Countries & 0.4595 \\
    Research areas & 0.3195 \\
    Papers Citations  & 0.3042 \\
    \bottomrule
    \end{tabular}%
  \label{tab:entropy}%
\end{table}%

An alternative measure of authorship concentration is Lotka's law. According to \cite{Lotka1926} empirical finding, authors' productivity follows a form of Zipf's law. The original finding, based on a restricted database of physics and chemistry, can be summarized by the equation:
\begin{equation}
a_n=\frac{a_1}{n^2}, \; n=1,2,\cdots, N
\label{eq:Lotka}
\end{equation}
where $a_n$ is the number of authors publishing $n$ papers, and $a_1$ is the number of authors publishing one paper. 
Power laws, as measures of concentration, are also used firms demography \citep{Zambrano2015}, population studies \citep{Hernando2012}, and other social applications \citep{Hernando2013}.

Considering that \cite{Lotka1926} deduced his empirical law from a very specific sample, a natural generalization could be:
\begin{equation}
a_n=\frac{a_1}{n^c}, \; n=1,2,\cdots, N
\label{eq:Lotkageneral}
\end{equation}
where $c$ is a parameter that should be estimated, so that it best fit data. 

In our sample $c=2.70$, with an $R^2=0.98$. Table \ref{tab:Lotka} summarizes the actual and fitted distribution of the number of authors publishing $n$ papers. We observe that the actual number of authors publishing only 1 paper is greater of what predicted by Lotka's law, which confirms that authorship is widely and more evenly distributed. Comparing our results to those by \cite{Chung1990}, we detect that author concentration in Bitcoin is lower than in several top financial journals (for any topic).

\begin{table}
  \centering
  \caption{Observed distribution of the number of authors who wrote a given number of papers, and Lotka's law fitted values}
    \begin{tabular}{rrrr}
    \toprule
    \# Articles & \# Authors & Observed frequency & Fitted frequency \\
    \midrule
    1     & 1885  & 0.8221 & 0.7081 \\
    2     & 262   & 0.1143 & 0.1087 \\
    3     & 67    & 0.0292 & 0.0363 \\
    4     & 26    & 0.0113 & 0.0167 \\
    5     & 20    & 0.0087 & 0.0091 \\
    6     & 12    & 0.0052 & 0.0056 \\
    7     & 9     & 0.0039 & 0.0037 \\
    8     & 6     & 0.0026 & 0.0026 \\
    9     & 2     & 0.0009 & 0.0019 \\
    10    & 3     & 0.0013 & 0.0014 \\
    17    & 1     & 0.0004 & 0.0003 \\
    \bottomrule
    \end{tabular}%
  \label{tab:Lotka}%
\end{table}%

%
\subsection{Citation, sources and authors graphs}
The following figures were generated using VOS viewer software, which allows to count the words which appear in the title, abstract and keywords to build all the relations which appear between different documents published in the Web of Science \citep{vanEck2010}.

Figure \ref{fig:keyword-all} represents the cloud map with relevant words of the article. This map shows how many times the words appear in the articles and how related are between them. The main finding is that the cloud could be divided into two parts. The right side is more related to economics and finance issues (blue and red) and the left side is more related to engineering and computer science (green and yellow). 
In the economics and finance part, we can distinguish between: (i) the blue part, more focused in the finance part of Bitcoin; and (ii) the red area, more specific in topics related to monetary economics. 
The engineering and computer science side also offers two distinctive subareas. One is more related more specifically to blockchain technology and smart contracts. The other is more related to mining protocols, security and cyber attacks. It is also relevant to observe that the expressions 'blockchain technology', 'money', and 'protocol' act as a nexus among different clusters.

\afterpage{
\begin{landscape}
\begin{figure}
    \centering
 \includegraphics[width=1.2\textwidth]{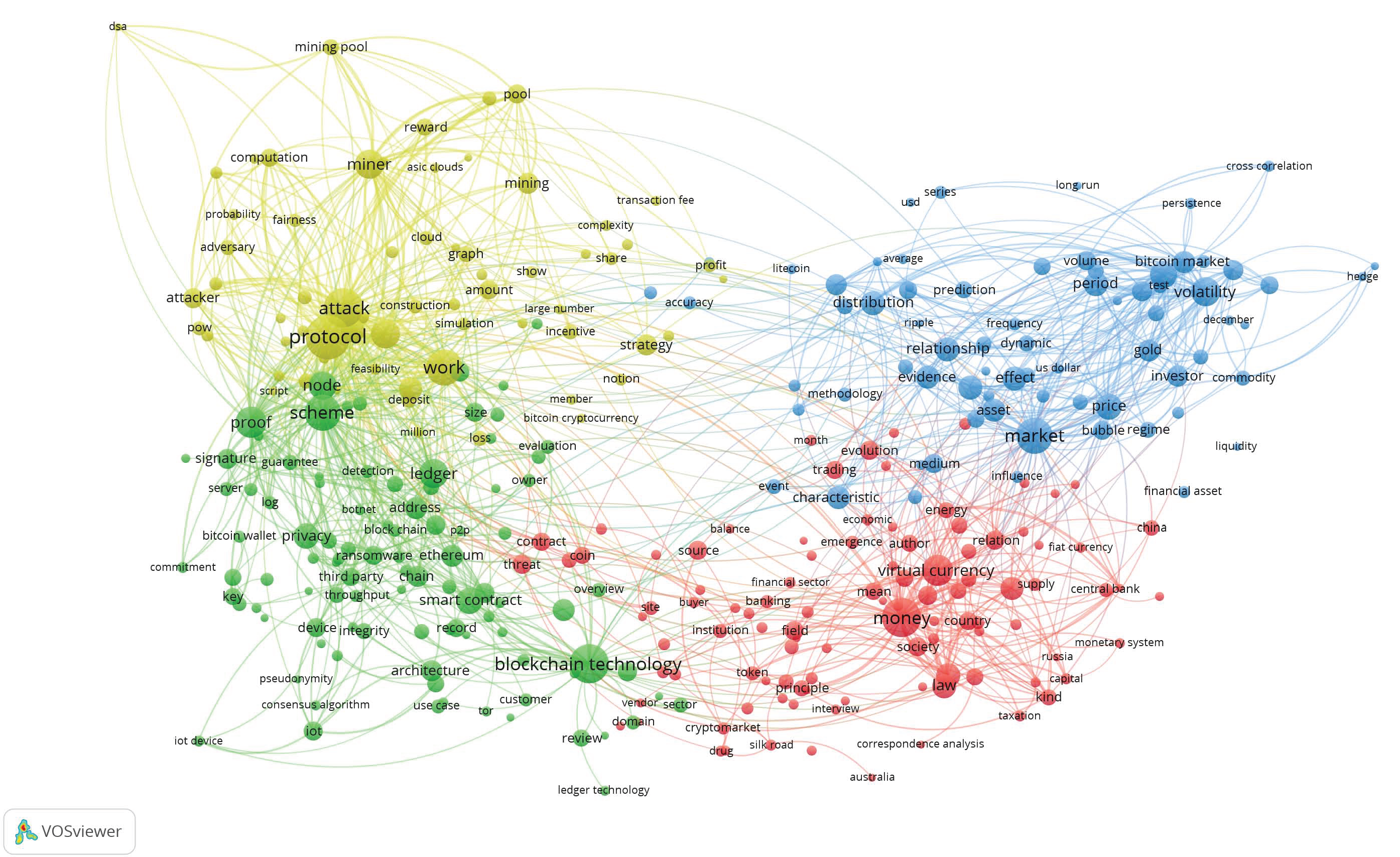}
\caption{Cloud map of words in titles and abstracts (full counting), generated with VOSviewer (\texttt{http://www.vosviewer.com/})}
\label{fig:keyword-all}
\end{figure}
\clearpage
\end{landscape}
}

Figure \ref{fig:keyword-binary} represents something similar to Figure \ref{fig:keyword-all}, with the slight difference that it is binary counted. This means that when a word appears, it is only counted once independently from how many times it appears in the document. This can change the results in the sense that if a word is very repeated in a document, it does not overestimate the results. In this cloud map, we can see that with the binary counting, there are three main clusters. The two economics and finance clusters merged into one bigger cluster. This difference suggests that the difference which appeared in Figure \ref{fig:keyword-all} could be because some specific words from economics or finance are enough repeated in the same article to create this difference.
The main keywords used are: blockchain and blockchain technology (blue), protocol (green), study and money (red). 

\afterpage{
\begin{landscape}
\begin{figure}
    \centering
          \includegraphics[width=1.2\textwidth]{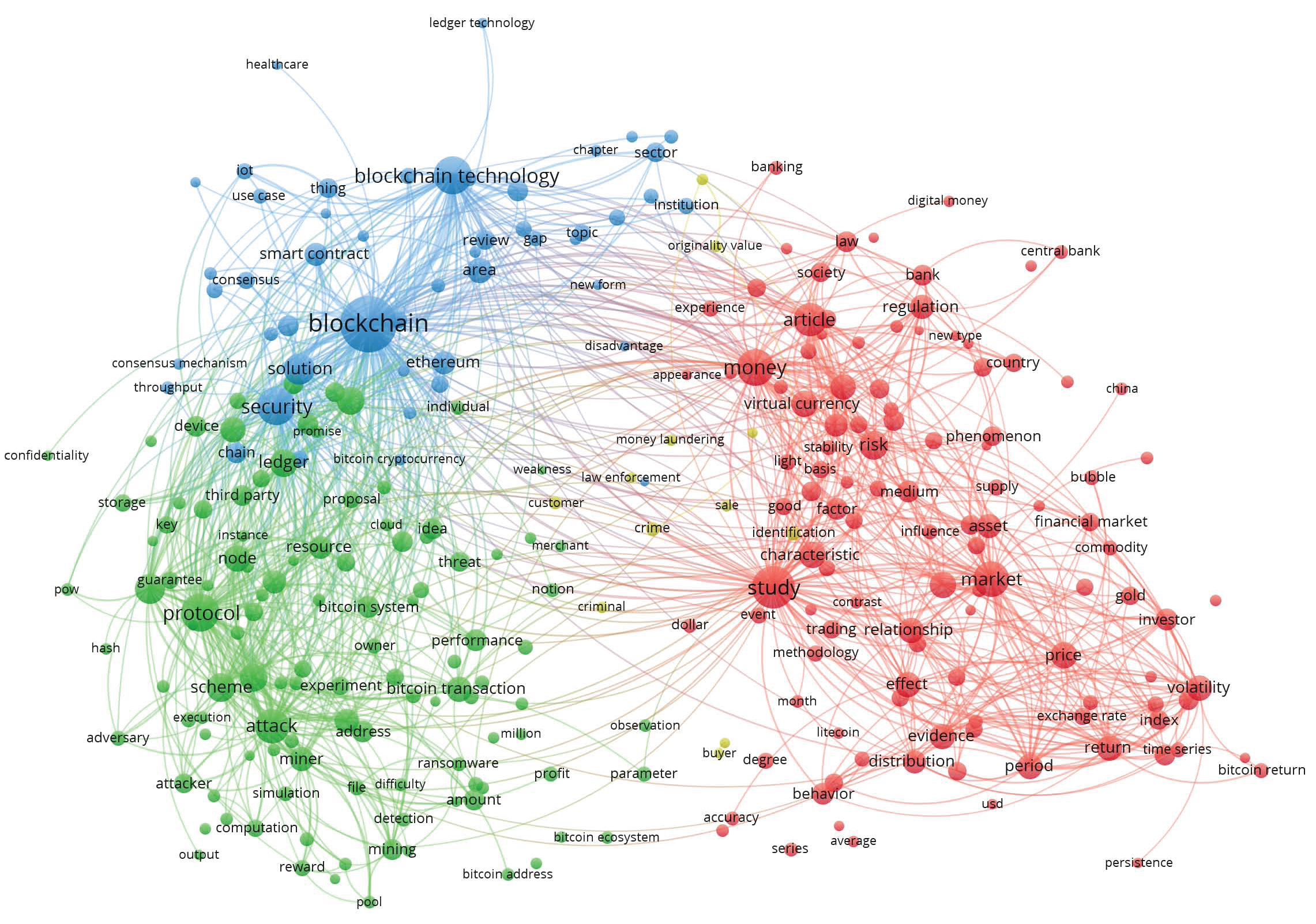}
\caption{Cloud map of words in titles and abstracts (binary counting), generated with VOSviewer (\texttt{http://www.vosviewer.com/})}
\label{fig:keyword-binary}
\end{figure}
\clearpage
\end{landscape}
}
In terms of the source of the articles, we observe in Figure \ref{fig:journals} again two well differentiated journals types: economics journals, and computer science and engineering journals. This result is consistent with those from Figure \ref{fig:keyword-all} and \ref{fig:keyword-binary}, in the sense that Bitcoin is a multidisciplinary field, where economists and computer scientists are the most interested in it. In agreement with Table \ref{tab:journals}, we detect several journals that are highly connected among them. In addition, \textit{PLOS ONE} occupies a position in the middle of the graph, which is expected as it is a multidisciplinary journal. We should also highlight that \textsl{Physica A} is located in the economics area. In spite of the fact that it is a statistical physics journal, it publishes many papers on quantitative finance and is a highly respected journal within the econophysics community. 
\afterpage{
\begin{landscape}
\begin{figure}
    \centering
          \includegraphics[width=1.2\textwidth]{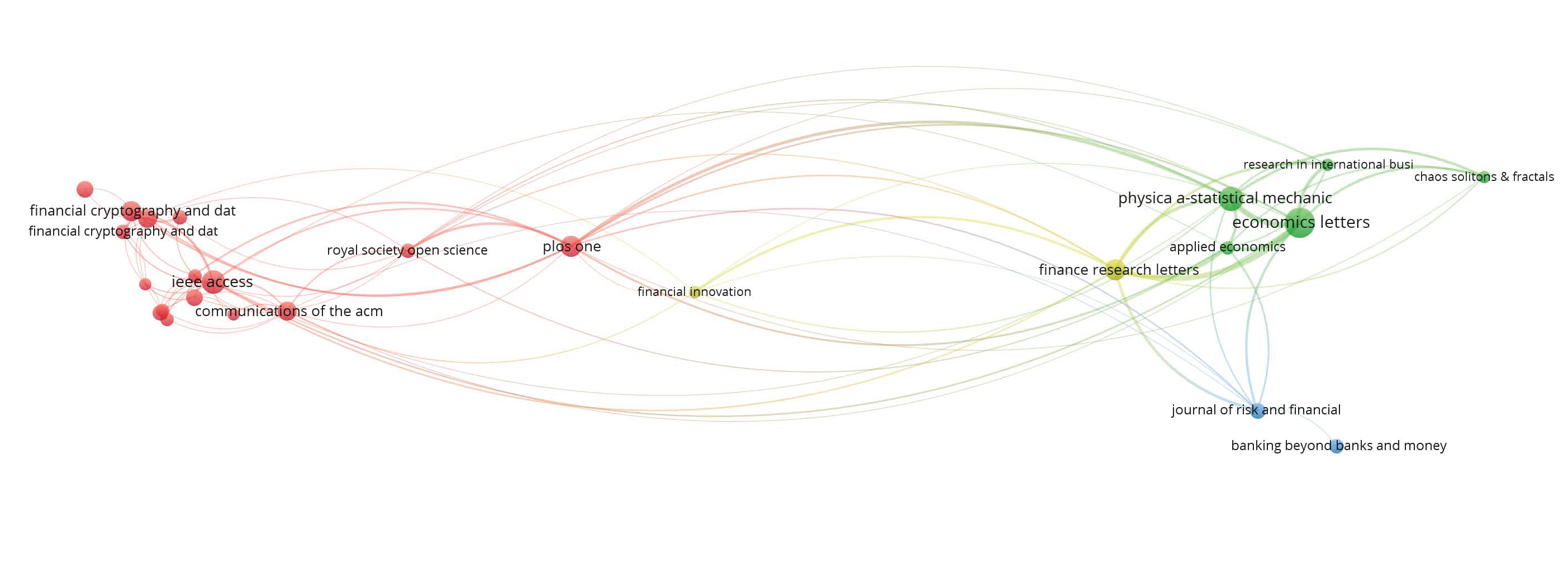}
\caption{Cloud map of journals where papers on Bitcoin were published, generated with VOSviewer (\texttt{http://www.vosviewer.com/})}
\label{fig:journals}
\end{figure}
\clearpage
\end{landscape}
}

Figure \ref{fig:authors} show all the articles and the size of the node depends on the number of citations. In this figure, we can observe that the most cited article from 2012 is the one written by \cite{Zyskind}, published in \textit{IEEE Security and Privacy Workshops (SPW)}, and dealing with the technological part of Bitcoin. The second one, which is the one written by \cite{Boehme2015} and published in the \textit{Journal of Economic Perspectives}, is more related to economics studies. There is a group on the right of the map which are closely related papers, mostly published in \textit{Economics Letters} and \textit{Finance Research Letters}, such as \cite{URQUHART201680}, \cite{CHEAH201532}, \cite{DYHRBERG201685} and \cite{Bariviera20171}.

There are some other outstanding articles in terms of number of citations such as the one by \cite{VANHOUT2014183} published in the \textit{International Journal of Drug Policy} or the one written by \cite{Miers2013} and published in the \textit{2013 IEEE Symposium On Security and Privacy (SP)}.
\afterpage{
\begin{landscape}
\begin{figure}
    \centering
          \includegraphics[width=1.2\textwidth]{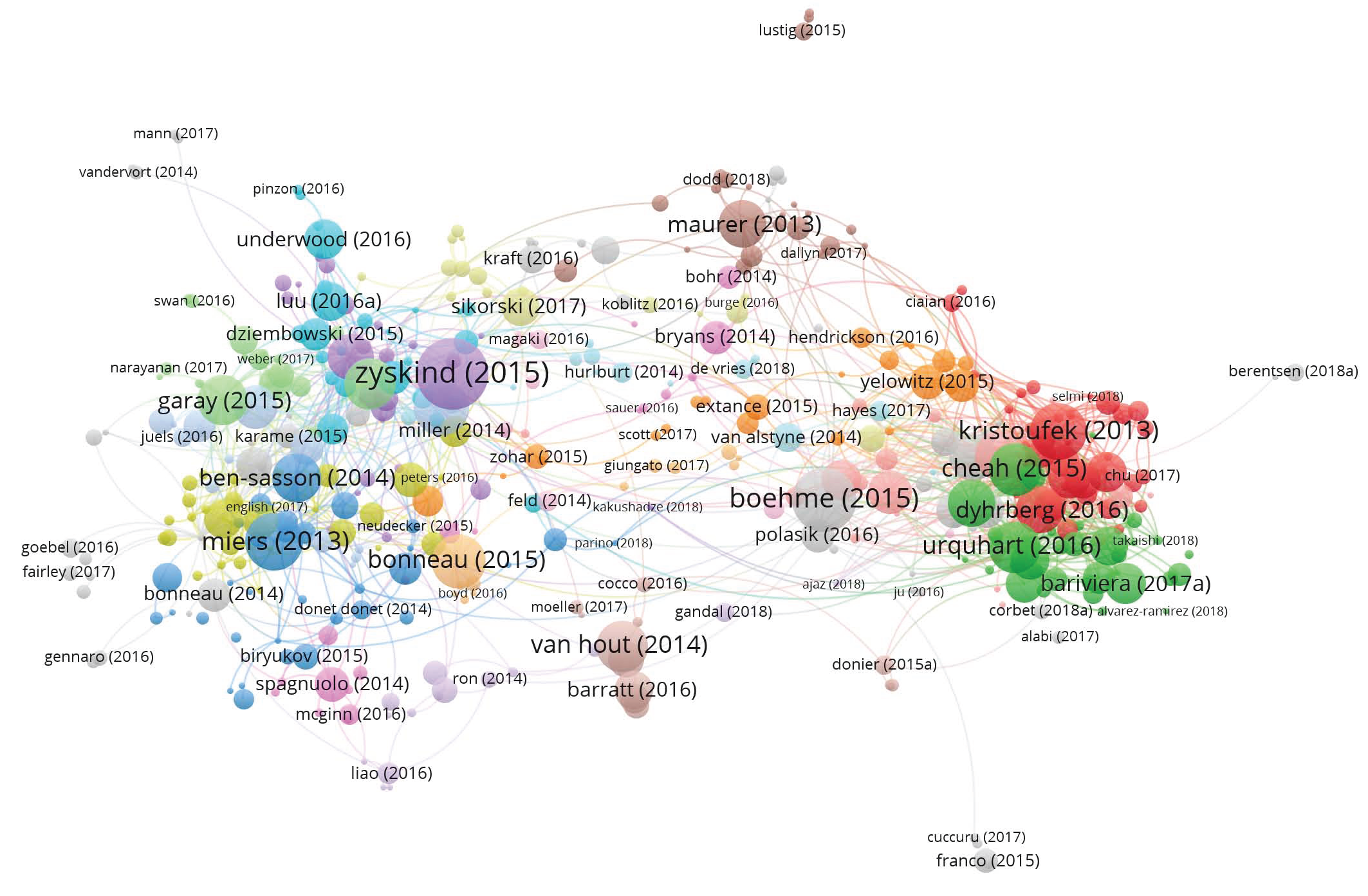}
\caption{Cloud map of journals of authors with papers on Bitcoin were published, generated with VOSviewer (\texttt{http://www.vosviewer.com/})}
\label{fig:authors}
\end{figure}
\clearpage
\end{landscape}
}

\section{Potential research lines}

Figure \ref{fig:keyword-binary} signalizes a clear separation between technology-oriented and economic-oriented papers. Considering the intrinsic technological component of bitcoin and its strong economic impact, we detect a lack of interdisciplinary works. For example, despite the correlation found among cryptocurrencies \citep{Zhang2018,Katsiampa2018,Klein2018,McMillan2019}, economic literature seems to overlook the impact of the technological setting (i.e. the different cryptocurrencies protocols) in the cross-behavior of cryptocurrencies. An exception is \cite{Aslanidis2019}, who speculate on the distinct Monero validation algorithm to justify its singular behavior. However, their explanation is provisional. Consequently, joint works by economists and computer scientists could add significant value to future research. 

There are several papers that find inefficiencies in bitcoin market. However, the reasons for such inefficiencies are unknown. Compared to traditional markets (bonds, stocks, etc.), where efficiency analysis has been tested against the behavior of institutional agents, levels or liquidity, macroeconomic shocks, among others, we detect a lack of such type of studies in the bitcoin market.

One of our findings is that legal studies of bitcoin are within the economic aspects. In this line, there is a lack of studies linking the economic impact of changes in legal regulations in bitcoin. 

Finally, research on behavioral aspects of bitcoin are needed, in order to understand a market, where there are heterogeneous individuals, interacting in real time. Related to this aspect, there are almost no paper regarding societal impact of bitcoin.

\section{Conclusions \label{sec:Conclusions}}

This study shows that cryptocurrencies' literature comprises mainly combination computer science and economics. Even though it is originally a technologically product, the foremost blockchain applications are the cryptocurrencies. Among them, Bitcoin is the dominant actor, both in the market and in the literature interest. 
The number of documents published on this topic has been increasing at a yearly rate of 124\%, albeit diminishing over the years. This paper constitutes the first comprehensive bibliometric study of Bitcoin literature, comprising all the papers indexed in the Web of Science since 2012. The large amount of data (1162 papers) allows to find significant results regarding top scholars, main journals and keywords of this multidisciplinary research field. We detected a high concentration in publishing countries. However, authors are diverse, and less concentrated than in leading finance journals. Additionally, citations are concentrated among a few papers.
In future works we would like to study the temporal evolution of keywords. Additionally, we would like to test if the quick growth and the subsequent fall in Bitcoin price in 2017 have affected the Bitcoin related research. Finally, related words such as ``cryptocurrencies'', ``fintech'', and ``peer-to-peer lending'' will be addressed in further research.

\bibliographystyle{apalike}
\bibliography{bibliometrics}

\begin{thebibliography}{}

\bibitem[Almunia et~al., 2009]{Almunia}
Almunia, M., B\'en\'etrix, A.~S., Eichengreen, B., O'Rourke, K.~H., and Rua, G.
  (2009).
\newblock {From Great Depression to Great Credit Crisis: Similarities,
  Differences and Lessons}.
\newblock NBER Working Papers 15524, National Bureau of Economic Research, Inc.

\bibitem[Andrikopoulos et~al., 2016]{Andrikopoulos201623}
Andrikopoulos, A., Samitas, A., and Kostaris, K. (2016).
\newblock Four decades of the journal of econometrics: Coauthorship patterns
  and networks.
\newblock {\em Journal of Econometrics}, 195(1):23 -- 32.

\bibitem[Aria and Cuccurullo, 2017]{bibliometrix}
Aria, M. and Cuccurullo, C. (2017).
\newblock bibliometrix: An r-tool for comprehensive science mapping analysis.
\newblock {\em Journal of Informetrics}, 11(4):959--975.

\bibitem[Aslanidis et~al., 2019]{Aslanidis2019}
Aslanidis, N., Bariviera, A.~F., and Martinez-Iba{\~{n}}ez, O. (2019).
\newblock An analysis of cryptocurrencies conditional cross correlations.
\newblock {\em Finance Research Letters}, 31:130--137.

\bibitem[Balcilar et~al., 2017]{BALCILAR201774}
Balcilar, M., Bouri, E., Gupta, R., and Roubaud, D. (2017).
\newblock Can volume predict bitcoin returns and volatility? a quantiles-based
  approach.
\newblock {\em Economic Modelling}, 64:74 -- 81.

\bibitem[Bariviera, 2017]{Bariviera20171}
Bariviera, A.~F. (2017).
\newblock {The inefficiency of Bitcoin revisited: A dynamic approach}.
\newblock {\em Economics Letters}, 161:1--4.

\bibitem[Bariviera et~al., 2018]{Bariviera2018}
Bariviera, A.~F., Zunino, L., and Rosso, O.~A. (2018).
\newblock {An analysis of high-frequency cryptocurrencies prices dynamics using
  permutation-information-theory quantifiers}.
\newblock {\em Chaos: An Interdisciplinary Journal of Nonlinear Science},
  28(7):075511.

\bibitem[Barro, 1979]{Barro1979}
Barro, R.~J. (1979).
\newblock Money and the price level under the gold standard.
\newblock {\em Economic Journal}, 89(353):13 -- 33.

\bibitem[B{\"{o}}hme et~al., 2015]{Bohme2015}
B{\"{o}}hme, R., Christin, N., Edelman, B., and Moore, T. (2015).
\newblock {Bitcoin: Economics, Technology, and Governance}.
\newblock {\em Journal of Economic Perspectives}, 29(2):213--238.

\bibitem[B\"ohme et~al., 2015]{Boehme2015}
B\"ohme, R., Christin, N., Edelman, B., and Moore, T. (2015).
\newblock Bitcoin: Economics, technology, and governance.
\newblock {\em Journal of Economic Perspectives}, 29(2):213--38.

\bibitem[Bonilla et~al., 2015]{Bonilla2015}
Bonilla, C.~A., Merig{\'{o}}, J.~M., and Torres-Abad, C. (2015).
\newblock {Economics in Latin America: a bibliometric analysis}.
\newblock {\em Scientometrics}, 105(2):1239--1252.

\bibitem[Bouri et~al., 2017]{BOURI2017192}
Bouri, E., Molnár, P., Azzi, G., Roubaud, D., and Hagfors, L.~I. (2017).
\newblock On the hedge and safe haven properties of bitcoin: Is it really more
  than a diversifier?
\newblock {\em Finance Research Letters}, 20:192 -- 198.

\bibitem[Broadus, 1987]{Broadus1987}
Broadus, R.~N. (1987).
\newblock Toward a definition of ``bibliometrics''.
\newblock {\em Scientometrics}, 12(5):373--379.

\bibitem[Cancino et~al., 2017]{Cancino2017614}
Cancino, C., Merig\'o, J.~M., Coronado, F., Dessouky, Y., and Dessouky, M.
  (2017).
\newblock Forty years of computers {\&} industrial engineering: A bibliometric
  analysis.
\newblock {\em Computers {\&} Industrial Engineering}, 113:614 -- 629.

\bibitem[Chatterjee et~al., 2018]{Chatterjee2018}
Chatterjee, J.~M., Son, L.~H., Ghatak, S., Kumar, R., and Khari, M. (2018).
\newblock Bitcoin exclusively informational money: a valuable review from 2010
  to 2017.
\newblock {\em Quality {\&} Quantity}, 52(5):2037--2054.

\bibitem[Cheah and Fry, 2015]{CHEAH201532}
Cheah, E.-T. and Fry, J. (2015).
\newblock Speculative bubbles in bitcoin markets? an empirical investigation
  into the fundamental value of bitcoin.
\newblock {\em Economics Letters}, 130:32 -- 36.

\bibitem[Chung and Cox, 1990]{Chung1990}
Chung, K.~H. and Cox, R. A.~K. (1990).
\newblock Patterns of productivity in the finance literature: A study of the
  bibliometric distributions.
\newblock {\em The Journal of Finance}, 45(1):301.

\bibitem[Ciaian et~al., 2016]{Ciaian2016}
Ciaian, P., Rajcaniova, M., and d’Artis Kancs (2016).
\newblock The economics of bitcoin price formation.
\newblock {\em Applied Economics}, 48(19):1799--1815.

\bibitem[{Clarivate Analytics}, 2018a]{indicatorhandbook}
{Clarivate Analytics} (2018a).
\newblock {\em {InCites Indicators Handbook}}.
\newblock Clarivate Analytics, Philadelphia (PA).
\newblock Accessed: 2019-02-20.

\bibitem[{Clarivate Analytics}, 2018b]{WoSinfo}
{Clarivate Analytics} (2018b).
\newblock {W}eb of {S}cience. trust the difference.
\newblock \url{http://wokinfo.com/}.
\newblock Accessed: 2019-02-18.

\bibitem[Claveau and Gingras, 2016]{Claveau2016}
Claveau, F. and Gingras, Y. (2016).
\newblock Macrodynamics of economics: A bibliometric history.
\newblock {\em History of Political Economy}, 4:551--592.

\bibitem[Cobo et~al., 2011]{Cobo2011}
Cobo, M., López-Herrera, A., Herrera-Viedma, E., and Herrera, F. (2011).
\newblock An approach for detecting, quantifying, and visualizing the evolution
  of a research field: A practical application to the fuzzy sets theory field.
\newblock {\em J. Informetrics}, 5:146--166.

\bibitem[Coinmarket, 2019]{coinmarketcap}
Coinmarket (2019).
\newblock {Crypto-Currency Market Capitalizations}.
\newblock \url{https://coinmarketcap.com/currencies/}.
\newblock Accessed: 2019-02-20.

\bibitem[Corbet et~al., 2019]{Corbet2019}
Corbet, S., Lucey, B., Urquhart, A., and Yarovaya, L. (2019).
\newblock {Cryptocurrencies as a financial asset: A systematic analysis}.
\newblock {\em International Review of Financial Analysis}, 62(June
  2018):182--199.

\bibitem[Corbet et~al., 2018]{CORBET201828}
Corbet, S., Meegan, A., Larkin, C., Lucey, B., and Yarovaya, L. (2018).
\newblock Exploring the dynamic relationships between cryptocurrencies and
  other financial assets.
\newblock {\em Economics Letters}, 165:28 -- 34.

\bibitem[Costa et~al., 2019]{Fonseca2019}
Costa, D.~F., Carvalho, F. d.~M., and Moreira, B. C. d.~M. (2019).
\newblock Behavioral economics and behavioral finance: A bibliometric analysis
  of the scientific fields.
\newblock {\em Journal of Economic Surveys}, 33(1):3--24.

\bibitem[Cronin, 2001]{Cronin2001}
Cronin, B. (2001).
\newblock {Hyperauthorship: A postmodern perversion or evidence of a structural
  shift in scholarly communication practices?}
\newblock {\em Journal of the American Society for Information Science and
  Technology}, 52(7):558--569.

\bibitem[Cuccurullo et~al., 2016]{CorradoScientometrix}
Cuccurullo, C., Aria, M., and Sarto, F. (2016).
\newblock Foundations and trends in performance management. a twenty-five years
  bibliometric analysis in business and public administration domains.
\newblock {\em Scientometrics}, 108(2):595--611.

\bibitem[Dabbagh et~al., 2019]{Dabbagh2019}
Dabbagh, M., Sookhak, M., and Safa, N.~S. (2019).
\newblock {The evolution of blockchain: A bibliometric study}.
\newblock {\em IEEE Access}, 7:19212--19221.

\bibitem[Dyhrberg, 2016a]{DYHRBERG201685}
Dyhrberg, A.~H. (2016a).
\newblock Bitcoin, gold and the dollar – a garch volatility analysis.
\newblock {\em Finance Research Letters}, 16:85 -- 92.

\bibitem[Dyhrberg, 2016b]{DYHRBERG2016139}
Dyhrberg, A.~H. (2016b).
\newblock Hedging capabilities of bitcoin. is it the virtual gold?
\newblock {\em Finance Research Letters}, 16:139 -- 144.

\bibitem[Fedor and Spellerberg, 2013]{Fedor2013}
Fedor, P. and Spellerberg, I. (2013).
\newblock {Shannon–Wiener Index}.
\newblock {\em Reference Module in Earth Systems and Environmental Sciences}.

\bibitem[Hart, 1971]{Hart1971}
Hart, P.~E. (1971).
\newblock {Entropy and Other Measures of Concentration}.
\newblock {\em Journal of the Royal Statistical Society.Series A (General)},
  134(1):pp. 73--85.

\bibitem[Hernando and Plastino, 2012]{Hernando2012}
Hernando, A. and Plastino, A. (2012).
\newblock Thermodynamics of urban population flows.
\newblock {\em Physical Review E - Statistical, Nonlinear, and Soft Matter
  Physics}, 86(6).

\bibitem[Hernando and Plastino, 2013]{Hernando2013}
Hernando, A. and Plastino, A. (2013).
\newblock Scale-invariance underlying the logistic equation and its social
  applications.
\newblock {\em Physics Letters A}, 377(3–4):176--180.

\bibitem[Holub and Johnson, 2018]{Holub2018}
Holub, M. and Johnson, J. (2018).
\newblock {Bitcoin research across disciplines}.
\newblock {\em Information Society}, 34(2):114--126.

\bibitem[Horowitz, 1970]{Horowitz1970}
Horowitz, I. (1970).
\newblock {Employment Concentration in the Common Market: An Entropy Approach}.
\newblock {\em Journal of the Royal Statistical Society.Series A (General)},
  133(3):pp. 463--479.

\bibitem[Huffman, 1952]{Huffman1952}
Huffman, D. (1952).
\newblock {A Method for the Construction of Minimum-Redundancy Codes}.
\newblock {\em Proceedings of the IRE}, 40(9):1098--1101.

\bibitem[Katsiampa, 2017]{KATSIAMPA20173}
Katsiampa, P. (2017).
\newblock Volatility estimation for bitcoin: A comparison of garch models.
\newblock {\em Economics Letters}, 158:3 -- 6.

\bibitem[Katsiampa, 2019]{Katsiampa2018}
Katsiampa, P. (2019).
\newblock {Volatility co-movement between Bitcoin and Ether}.
\newblock {\em Finance Research Letters}, forthcoming.

\bibitem[Khan and Salah, 2018]{KHAN2018395}
Khan, M.~A. and Salah, K. (2018).
\newblock Iot security: Review, blockchain solutions, and open challenges.
\newblock {\em Future Generation Computer Systems}, 82:395 -- 411.

\bibitem[Klein et~al., 2018]{Klein2018}
Klein, T., {Pham Thu}, H., and Walther, T. (2018).
\newblock {Bitcoin is not the New Gold – A comparison of volatility,
  correlation, and portfolio performance}.
\newblock {\em International Review of Financial Analysis}, 59:105--116.

\bibitem[Korom, 2019]{Korom2019}
Korom, P. (2019).
\newblock A bibliometric visualization of the economics and sociology of wealth
  inequality: a world apart?
\newblock {\em Scientometrics}.

\bibitem[Kuo et~al., 2017]{Kuo2017}
Kuo, T.-T., Kim, H.-E., and Ohno-Machado, L. (2017).
\newblock {Blockchain distributed ledger technologies for biomedical and health
  care applications}.
\newblock {\em Journal of the American Medical Informatics Association},
  24(6):1211--1220.

\bibitem[Lahmiri and Bekiros, 2018]{LAHMIRI201828}
Lahmiri, S. and Bekiros, S. (2018).
\newblock Chaos, randomness and multi-fractality in bitcoin market.
\newblock {\em Chaos, Solitons \& Fractals}, 106:28 -- 34.

\bibitem[Lamberti et~al., 2004]{Lamberti2004119}
Lamberti, P.~W., Mart\'{\i}n, M.~T., Plastino, A., and Rosso, O.~A. (2004).
\newblock Intensive entropic non-triviality measure.
\newblock {\em Physica A}, 334(1–2):119--131.

\bibitem[Liu, 2016]{Liu2016}
Liu, J. (2016).
\newblock {Bitcoin Literature: a Co-Word Analysis}.
\newblock In Cermakova, K., editor, {\em 6th Economics {\&} Finance Conference,
  OECD, Paris}, number September, pages 262--272, Paris. International
  Institute of Social and Economic Sciences.

\bibitem[Liu, 2018]{Liu2018}
Liu, W. (2018).
\newblock Portfolio diversification across cryptocurrencies.
\newblock {\em Finance Research Letters}, (July).

\bibitem[Lotka, 1926]{Lotka1926}
Lotka, A.~J. (1926).
\newblock The frequency distribution of scientific productivity.

\bibitem[McMillan, 2019]{McMillan2019}
McMillan, D.~G. (2019).
\newblock {Cross-asset relations, correlations and economic implications}.
\newblock {\em Global Finance Journal}.

\bibitem[Meng et~al., 2018]{Meng}
Meng, W., Tischhauser, E.~W., Wang, Q., Wang, Y., and Han, J. (2018).
\newblock When intrusion detection meets blockchain technology: A review.
\newblock {\em IEEE Access}, 6:10179--10188.

\bibitem[Merig{\'o} et~al., 2016]{Merigo2016}
Merig{\'o}, J.~M., Cancino, C.~A., Coronado, F., and Urbano, D. (2016).
\newblock Academic research in innovation: a country analysis.
\newblock {\em Scientometrics}, 108(2):559--593.

\bibitem[Miau and Yang, 2018]{Miau2018}
Miau, S. and Yang, J.-M. (2018).
\newblock {Bibliometrics-based evaluation of the Blockchain research trend:
  2008-March 2017}.
\newblock {\em Technology Analysis {\&} Strategic Management},
  30(9):1029--1045.

\bibitem[Miers et~al., 2013]{Miers2013}
Miers, I., Garman, C., Green, M., and Rubin, A.~D. (2013).
\newblock Zerocoin: Anonymous distributed e-cash from bitcoin.
\newblock In {\em 2013 IEEE Symposium on Security and Privacy}, pages 397--411.

\bibitem[Nadarajah and Chu, 2017]{NADARAJAH20176}
Nadarajah, S. and Chu, J. (2017).
\newblock On the inefficiency of bitcoin.
\newblock {\em Economics Letters}, 150:6 -- 9.

\bibitem[Nakamoto, 2009]{Nakamoto}
Nakamoto, S. (2009).
\newblock Bitcoin: A peer-to-peer electronic cash system.
\newblock \url{https://bitcoin.org/bitcoin.pdf/}.
\newblock Accessed: 2016-12-27.

\bibitem[Petersen et~al., 2008]{Petersen2008}
Petersen, K., Feldt, R., Mujtaba, S., and Mattsson, M. (2008).
\newblock Systematic mapping studies in software engineering.
\newblock In {\em Proceedings of the 12th International Conference on
  Evaluation and Assessment in Software Engineering}, EASE'08, pages 68--77,
  Swindon, UK. BCS Learning \& Development Ltd.

\bibitem[Pielou, 1966]{Pielou1966}
Pielou, E. (1966).
\newblock {The measurement of diversity in different types of biological
  collections}.
\newblock {\em Journal of Theoretical Biology}, 13:131--144.

\bibitem[Platanakis and Urquhart, 2019]{PLATANAKIS2019}
Platanakis, E. and Urquhart, A. (2019).
\newblock {Portfolio management with cryptocurrencies: The role of estimation
  risk}.
\newblock {\em Economics Letters}, 177:76--80.

\bibitem[Polyakov et~al., 2017]{Polyakov2017}
Polyakov, M., Polyakov, S., and Iftekhar, M.~S. (2017).
\newblock {Does academic collaboration equally benefit impact of research
  across topics? The case of agricultural, resource, environmental and
  ecological economics}.
\newblock {\em Scientometrics}, 113(3):1385--1405.

\bibitem[Pritchard, 1969]{Pritchard1969}
Pritchard, A. (1969).
\newblock Statistical bibliography or bibliometrics?
\newblock {\em Journal of Documentation}, 25:348--349.

\bibitem[Rosso et~al., 2007]{Rosso07}
Rosso, O.~A., Larrondo, H.~A., Martin, M.~T., Plastino, A., and Fuentes, M.~A.
  (2007).
\newblock Distinguishing noise from chaos.
\newblock {\em Phys. Rev. Lett.}, 99(15):154102.

\bibitem[Shahzad et~al., 2019]{Shahzad2019}
Shahzad, S. J.~H., Bouri, E., Roubaud, D., Kristoufek, L., and Lucey, B.
  (2019).
\newblock {Is Bitcoin a better safe-haven investment than gold and
  commodities?}
\newblock {\em International Review of Financial Analysis}, 63:322--330.

\bibitem[Shannon and Weaver, 1949]{book:shannon1949}
Shannon, C.~E. and Weaver, W. (1949).
\newblock {\em The Mathematical Theory of Communication}.
\newblock University of Illinois Press, Champaign, IL.

\bibitem[Smales, 2018]{Smales2018}
Smales, L. (2018).
\newblock {Bitcoin as a safe haven: Is it even worth considering?}
\newblock {\em Finance Research Letters}.

\bibitem[Urquhart, 2016]{URQUHART201680}
Urquhart, A. (2016).
\newblock The inefficiency of bitcoin.
\newblock {\em Economics Letters}, 148:80 -- 82.

\bibitem[Urquhart, 2017]{URQUHART2017145}
Urquhart, A. (2017).
\newblock Price clustering in bitcoin.
\newblock {\em Economics Letters}, 159:145 -- 148.

\bibitem[van Eck and Waltman, 2010]{vanEck2010}
van Eck, N.~J. and Waltman, L. (2010).
\newblock Software survey: Vosviewer, a computer program for bibliometric
  mapping.
\newblock {\em Scientometrics}, 84(2):523--538.

\bibitem[{Van Hout} and Bingham, 2014]{VANHOUT2014183}
{Van Hout}, M.~C. and Bingham, T. (2014).
\newblock Responsible vendors, intelligent consumers: Silk road, the online
  revolution in drug trading.
\newblock {\em International Journal of Drug Policy}, 25(2):183 -- 189.

\bibitem[Wei, 2019]{Wei2019}
Wei, G. (2019).
\newblock A bibliometric analysis of the top five economics journals during
  2012–2016.
\newblock {\em Journal of Economic Surveys}, 33(1):25--59.

\bibitem[Yermack, 2013]{NBERw19747}
Yermack, D. (2013).
\newblock Is bitcoin a real currency? an economic appraisal.
\newblock Working Paper 19747, National Bureau of Economic Research.

\bibitem[Yli-Huumo et~al., 2016]{Yli-Huumo2016}
Yli-Huumo, J., Ko, D., Choi, S., Park, S., and Smolander, K. (2016).
\newblock {Where Is Current Research on Blockchain Technology?-A Systematic
  Review.}
\newblock {\em PloS one}, 11(10):e0163477.

\bibitem[Zambrano et~al., 2015]{Zambrano2015}
Zambrano, E., Hernando, A., {Fern{\'{a}}ndez Bariviera}, A., Hernando, R., and
  Plastino, A. (2015).
\newblock {Thermodynamics of firms' growth}.
\newblock {\em Journal of The Royal Society Interface}, 12(112):20150789.

\bibitem[Zeng et~al., 2018]{Zeng2018}
Zeng, S., Ni, X., Yuan, Y., and Wang, F.-Y. (2018).
\newblock A bibliometric analysis of blockchain research.
\newblock In {\em 2018 IEEE Intelligent Vehicles Symposium (IV)}, pages
  102--107.

\bibitem[Zhang et~al., 2018]{Zhang2018}
Zhang, W., Wang, P., Li, X., and Shen, D. (2018).
\newblock {The inefficiency of cryptocurrency and its cross-correlation with
  Dow Jones Industrial Average}.
\newblock {\em Physica A: Statistical Mechanics and its Applications},
  510:658--670.

\bibitem[Zheng et~al., 2018]{Zheng2018}
Zheng, Z., Xie, S., Dai, H.-N., Chen, X., and Wang, H. (2018).
\newblock Blockchain challenges and opportunities: a survey.
\newblock {\em International Journal of Web and Grid Services}, 14(4):352--375.

\bibitem[Zyskind et~al., 2015]{Zyskind}
Zyskind, G., Nathan, O., and Pentland, A.~S. (2015).
\newblock Decentralizing privacy: Using blockchain to protect personal data.
\newblock In {\em 2015 IEEE Security and Privacy Workshops}, pages 180--184.

\end{thebibliography}

\end{document}